\documentclass[prl,twocolumn,showpacs]{revtex4}
\usepackage{amsfonts}
\usepackage{amsmath}
\usepackage{amssymb}
\usepackage{graphicx}

\providecommand{\U}[1]{\protect\rule{.1in}{.1in}}

\begin{document}

\title{Entanglement Dynamics in Harmonic Oscillator Chains}
\author{R. G. Unanyan and M. Fleischhauer}
\affiliation{Fachbereich Physik und Forschungszentrum OPTIMAS, Technische Universit\"{a}t
Kaiserslautern, D-67663 Kaiserslautern, Germany}
\date{\today}

\begin{abstract}
We study the long-time evolution of the bipartite entanglement in
translationally invariant 1D harmonic lattice systems. We show that for a wide
class of Hamiltonians and generic initial states there exists a \textit{lower%
} bound for the von Neumann entropy which increases linearly in time. This
implies that the dynamics of harmonic lattice systems can in general not
efficiently be simulated by algorithms based on matrix-product
decompositions of the quantum state.
\end{abstract}

\maketitle

Recently, so-called matrix product states (MPS) have received much interest
for the numerical simulation of one-dimensional quantum many body systems
\cite{Vidal2004},\cite{Schuch}. This is because the ground state of
fermionic and bosonic lattice systems with finite-range interactions and an
excitation gap usually implies an area law of entanglement \cite{Eisert},
stating that the von Neumann entropy of a partition scales with the surface
size. In 1D the surface area is independent on the size of the system
resulting in a weakly entangled ground state, which can thus faithfully be
represented by MPS. Even when the excitation gap vanishes, i.e. for critical
systems, there is only a correction which is at most logarithmic in the
system size. The situation is however quite different for non-equilibrium
problems as here not only the scaling with size but also with time is
relevant. With respect to the latter only an upper bound derived by Lieb and
Robinson exists \cite{Lieb-Robinson}, which states that the von Neumann
entropy increases at most linear in time. Being an upper bound, it does of
course not allow to draw any conclusion about the approximability of the
long-time dynamics of quantum many-body systems by MPS. However, it is very
often found, that the bi-partite entropy does indeed scale linear in time,
which implies that the required computational resources increase exponentially in time. For
example it has been shown for the spin-$\frac{1}{2}$ $XY$ model that the
entropy grows linearly with time after a global quench \cite{Calabrese}. On
the other hand, it was found recently for the case of free fermions that the
entropy can grow only logarithmically in time \cite{Klich}, showing that for
certain initial states the long-time dynamics is accessible with MPS
based methods. If for free fermions the scaling of the entanglement entropy
in time is only moderate for certain initial conditions, what kind of
scaling with time can we expect for the entanglement entropy of free bosons?

In the present paper we study the time evolution of the entropy in 1D
bosonic systems that evolve under translationally invariant quadratic
Hamiltonians with local or finite-range couplings. In order to separate the
problem of size-scaling from the scaling with time, which is the subject
of interest in this paper, we consider a specific
class of 1D systems where the bi-partite entropy becomes independent on
system size. To this end we choose as initial state of the time evolution the
ground state $\Phi $ of some local, gapped Hamiltonian $H_{0}$. As explained
in the following, one can intuitively expect that under such initial 
conditions the bi-partite entanglement entropy of the time-evolved state 
will be independent on system size. Since for the ground state of local 
Hamiltonians the presence of an excitation gap is sufficient for an area 
law of entanglement \cite{Eisert},
the entropy of the initial state is size independent. It is easy to see that
the state time-evolved under a Hamiltonian $H$, $\Psi (t)=\exp \{-iHt\}\Phi $, 
is the ground state of a time-dependent Hamiltonian $H^{\prime }[t]=\exp \{-iHt\}H_{0}\exp \{iHt\}$.
The spectrum of $H^{\prime }[t]$ is identical to that of $H_{0}$, i.e. it
too has an excitation gap. Furthermore the Lieb-Robinson bounds guarantee that
for any fixed time $t$ its coupling matrix elements between sites $i$ and $j$
are exponentially small beyond a certain distance $l_{c}$, i.e. for $%
|i-j|>l_{c}$. Thus $H^{\prime }[t]$ is also of finite range \cite{footnote}.
As a consequence we can expect that the entanglement entropy of the time-evolved state
will saturate with increasing system size.
However it is not possible to draw any conclusion about the time-scaling of
entanglement beyond the limits set by the Lieb-Robinson {\it upper} bounds.

In the present paper we show that in contrast to free fermions the entropy
of the time evolved quantum state \emph{always} grows linearly in time,
making a long-time simulation of 1D bosonic systems with MPS based methods
impossible.


\textit{Translationally invariant harmonic oscillators:} To be specific we
consider a one-dimensional system of $N$ bosonic oscillators described by $N$
pairs of canonical operators $\mathbf{x}=\left( x_{1},x_{2},...,x_{N}\right)
$ and $\mathbf{p}=\left( p_{1},p_{2},...,p_{N}\right) $. The oscillators are
coupled by a quadratic Hamiltonian of the form
\begin{equation}
H=\frac{1}{2}\mathbf{p}^{2}+\frac{1}{2}\left\langle \mathbf{x}\right\vert
V\left\vert \mathbf{x}\right\rangle ,  \label{Hamiltonian}
\end{equation}
where $V$ is a real, symmetric, positive definite, time-independent matrix.
We assume translational invariance, implying that $V$ is a Toeplitz matrix.
Furthermore we consider periodic boundary conditions, such that $V$ is
circulant. Circulant matrices form a commutative algebra. Moreover the
elements of a circulant matrix can be generated from the spectral function $%
\lambda (\theta )$, i.e.
\begin{equation}
V_{kl}=\frac{1}{2\pi }\int_{0}^{2\pi }\!\!\mathrm{d}\theta \,\lambda (\theta
)\,e^{-i(k-l)\theta }.
\end{equation}

We now want to determine the scaling of the entanglement entropy with time.
To this end we have to find the time evolution under the local Hamiltonian (%
\ref{Hamiltonian}). The system is assumed to start its evolution at $t=0$
from a Gaussian state, i.e.
\begin{equation}
\Phi \left( \mathbf{x}\right) =\alpha _{0}\exp \left( -\frac{1}{2}%
\left\langle \mathbf{x}\right\vert B\left\vert \mathbf{x}\right\rangle
\right) ,\text{ }\alpha _{0}=\left( \frac{\det B}{\pi ^{N}}\right) ^{1/4}
\label{initialcondition}
\end{equation}
where $B$ is a real, symmetric, and positive definite Toeplitz matrix.
Periodic boundary conditions imply that $B$ is also a circulant matrix with
spectral function $\beta (\theta )$. As the initial state is the ground
state of a gapped, local Hamiltonian, $\beta (\theta )$ is non-zero and
regular corresponding to a non-critical state. Since the Hamiltonian of the
system is quadratic, the time-evolved state remains Gaussian and we have to
search for a solution in the form
\begin{equation}
\Psi \left( \mathbf{x},t\right) =\frac{1}{\left( \pi ^{N}\det \widetilde{A}%
^{-1}\right) ^{1/4}}\exp \left( -\frac{1}{2}\left\langle \mathbf{x}%
\right\vert A\left( t\right) \left\vert \mathbf{x}\right\rangle \right) .
\label{StateEvolution}
\end{equation}
Here and in the following a tilde denotes the real part, i.e. $\widetilde{X}$
$=\frac{X+X^{\ast }}{2}$. By taking into account the symmetry of $B$ and
after simple calculations one can easily find that $A\left( t\right) $ obeys
the Riccati equation
\begin{equation}
i\frac{\partial A}{\partial t}=A^{2}-V,\qquad A\left( 0\right) =B.
\label{A_equation}
\end{equation}
Its solution can be written as
\begin{equation}
A\left( t\right) =V^{1/2}\frac{\cos \left( tV^{1/2}\right) B+iV^{1/2}\sin
\left( tV^{1/2}\right) }{\cos \left( tV^{1/2}\right) V^{1/2}+i\sin \left(
tV^{1/2}\right) B}  \label{ASolution}
\end{equation}
which is again a circulant matrix. The spectral function $\Lambda (\theta
,t) $ of its real part $\widetilde{A}$ can easily be obtained from $\lambda
(\theta )$ and $\beta (\theta )$:
\begin{equation}
\Lambda \left( \theta ,t\right) =\frac{\beta \left( \theta \right) \lambda
\left( \theta \right) }{\lambda \left( \theta \right) \cos ^{2}\left(
t\lambda ^{1/2}\left( \theta \right) \right) +\beta ^{2}\left( \theta
\right) \sin ^{2}\left( t\lambda ^{1/2}\left( \theta \right) \right) },
\label{symbolfunction}
\end{equation}
Note that if $B=V^{1/2}$, the spectral function and thus the matrix $A(t)$
becomes time-independent, as in this case the initial state is the ground
state of the full Hamiltonian.


\textit{Reduced density matrix:} Having the solution of the Schr\"{o}dinger
equation we can now calculate the reduced density matrix of a block of $N-n$
oscillators. The calculations can be done by partitioning the symmetric
matrices $A\left( t\right) $ and ${A}^{-1}\left( t\right) $ into blocks
\begin{equation}
A\left( t\right) =\left[
\begin{array}{cc}
T & C \\
C^{T} & R%
\end{array}%
\right] ,\text{ \ }{A}^{-1}\left( t\right) =\left[
\begin{array}{cc}
Q & D \\
D^{T} & P%
\end{array}%
\right]  \label{Apartition}
\end{equation}%
where $T$ is an $n\times n$ and $R$ an $\left( N-n\right) \times \left(
N-n\right) $ matrix. Similar calculations have been done in \cite{Bombelli}
and \cite{Srednickii} for the ground state of a chain of oscillators. \
After a lengthy but straightforward calculation we find for the matrix elements of the reduced density operator
\begin{equation}
\rho _{R}\left( \mathbf{x},\mathbf{x}^{\prime }\right) =\mathcal{N}\exp %
\left[ \left(
\begin{array}{c}
\mathbf{x} \\
\mathbf{x^{\prime }}%
\end{array}%
\right) ^{T}\left[
\begin{array}{cc}
-\Gamma & \Delta \\
\Delta ^{\ast } & -\Gamma ^{\ast }%
\end{array}%
\right] \left(
\begin{array}{c}
\mathbf{x} \\
\mathbf{x^{\prime }}%
\end{array}%
\right) \right] ,  \label{reducedDensityMatrix}
\end{equation}%
where $\mathbf{x}=(x_{n+1},....x_{N}),$ $\mathbf{x^{\prime }}%
=(x_{n+1}^{\prime },....x_{N}^{\prime })$ are the coordinates of the
remaining $N-n$ oscillators,
\begin{equation*}
\Gamma =\frac{R}{2}-\frac{C^{T}\widetilde{T}^{-1}C}{4},\text{ }\Delta =\frac{%
C^{T}\widetilde{T}^{-1}C^{\ast }}{4},
\end{equation*}%
and $\mathcal{N}=\left( \det {\widetilde{P}}^{-1}\right) ^{1/2}/{\left( \pi
\right) ^{\frac{N-n}{2}}}$ is a normalization with
\begin{equation}
{\widetilde{P}}^{-1}=\widetilde{R}-\widetilde{C}^{T}\widetilde{T}^{-1}%
\widetilde{C},  \label{momentumcorrelation}
\end{equation}%
Here and in the following $\widetilde{Y}^{-1}$ denotes $\left( \widetilde{Y}%
\right) ^{-1}$.


\textit{Purity and lower bound for the Entropy: } We proceed by analyzing
the dynamical behavior of the bi-partite entanglement. There are several
measures of entanglement between parties of a closed system, examples being
the von Neumann entropy $S=-\mathrm{tr}\left( \rho _{R}\ln \rho _{R}\right) $
\ and the purity $\mathrm{tr}\rho _{R}^{2}$ , where the following inequality
holds $S\geq -\ln \mathrm{tr}[\rho _{R}^{2}]$. It should be noted that $-%
\mathrm{\ln tr}\rho _{R}^{2}$ represents also a lower bound to all Renyi
entropies $S_{\alpha }=\frac{1}{1-\alpha }\ln \mathrm{tr}\left[ \rho
_{R}^{\alpha }\right] $ with $\alpha <1$ as $S_{\alpha }>S_{1}=S$.

In order to derive a lower bound for the entropy we calculate the purity of (%
\ref{reducedDensityMatrix}).
\begin{equation}
\mathrm{tr}\left[ \rho _{R}^{2}\right] =\int \mathrm{d}\mathbf{x}\mathrm{d}%
\mathbf{x^{\prime }}\rho _{R}\left( \mathbf{x},\mathbf{x^{\prime }}\right)
\rho _{R}\left( \mathbf{x^{\prime }},\mathbf{x}\right) .  \label{Purity}
\end{equation}%
The Gaussian nature of (\ref{reducedDensityMatrix}) allows to calculate this
integral in a straight-forward way:
\begin{equation}
\mathrm{tr}\left[ \rho _{R}^{2}\right] =\frac{\left( \det {\widetilde{P}}%
^{-1}\right) }{\left( \det \left[ \widetilde{\Gamma }-\widetilde{\Delta }%
\right] \det \left[ \widetilde{\Gamma }+\widetilde{\Delta }\right] \right)
^{1/2}}.  \label{purityExpress}
\end{equation}%
After simple algebra one obtains
\begin{equation}
\mathrm{tr}\left[ \rho _{R}^{2}\right] =\left[ \det \left( \widetilde{P}%
\left( \widetilde{R}+Z^{T}\widetilde{T}^{-1}Z\right) \right) \right]
^{-1/2}\!\!\leq \left[ \det \left( \widetilde{P}\widetilde{R}\right) \right]
^{-1/2},  \notag
\end{equation}%
where $Z=(C-C^{\ast })/{2i}$. The last inequality follows from the fact that
$Z^{T}\widetilde{T}^{-1}Z$ is a positive definite matrix. With this we find
the following lower bound to the von-Neumann entropy
\begin{equation}
S\geq \frac{1}{2}\ln \det \left( \ \widetilde{P}\cdot \widetilde{R}\right) .
\label{EntropyLowerBound}
\end{equation}%
In order to facilitate analytical calculations of determinants, we consider
the limits $N\gg 1$ and $N>n\gg 1.$It can then be shown \cite{Unanyan2005}
that in this limit the elements of matrices $\widetilde{R}$ and $\widetilde{P%
}$ can be generated from the spectral functions $\Lambda \left( \theta
,t\right) $ and $\Lambda ^{-1}\left( \theta ,t\right) $ respectively. As $%
\Lambda \left( \theta ,t\right) $ is a regular function i.e. $\Lambda \left(
\theta ,t\right) >0$ for any $t$, we may apply the strong Szeg\"{o} theorem
\cite{Szego} to calculate the determinants. According to this theorem
\begin{equation}
S\geq {\displaystyle\sum\limits_{k=1}^{\infty }}k\left\vert c_{k}\right\vert
^{2},  \label{LowerBoundFourier}
\end{equation}%
where the $c_{k}$ are Fourier coefficients of $\ln \Lambda ^{-1}\left(
\theta ,t\right) $, i.e.,
\begin{equation}
c_{k}=\frac{1}{2\pi }{\displaystyle\int\limits_{0}^{2\pi }}d\theta \,\ln
\Lambda ^{-1}\left( \theta ,t\right) \exp \left( -i\theta k\right) .
\label{Fourier}
\end{equation}%
If $\beta \left( \theta \right) $\ and $\lambda (\theta )$\ are constant
function i.e. the oscillators are uncoupled, all Fourier coefficents (\ref%
{Fourier}) vanish except for $c_{0}$. In this case (\ref{LowerBoundFourier})
reduces to the trivial bound $S\geq 0$\ ( entanglement is never generated).
In what follows we will consider only the non trivial case when $\lambda
(\theta )$\ is a not constant.

\begin{figure}[tbp]
\centering
\includegraphics[width=0.95\columnwidth]{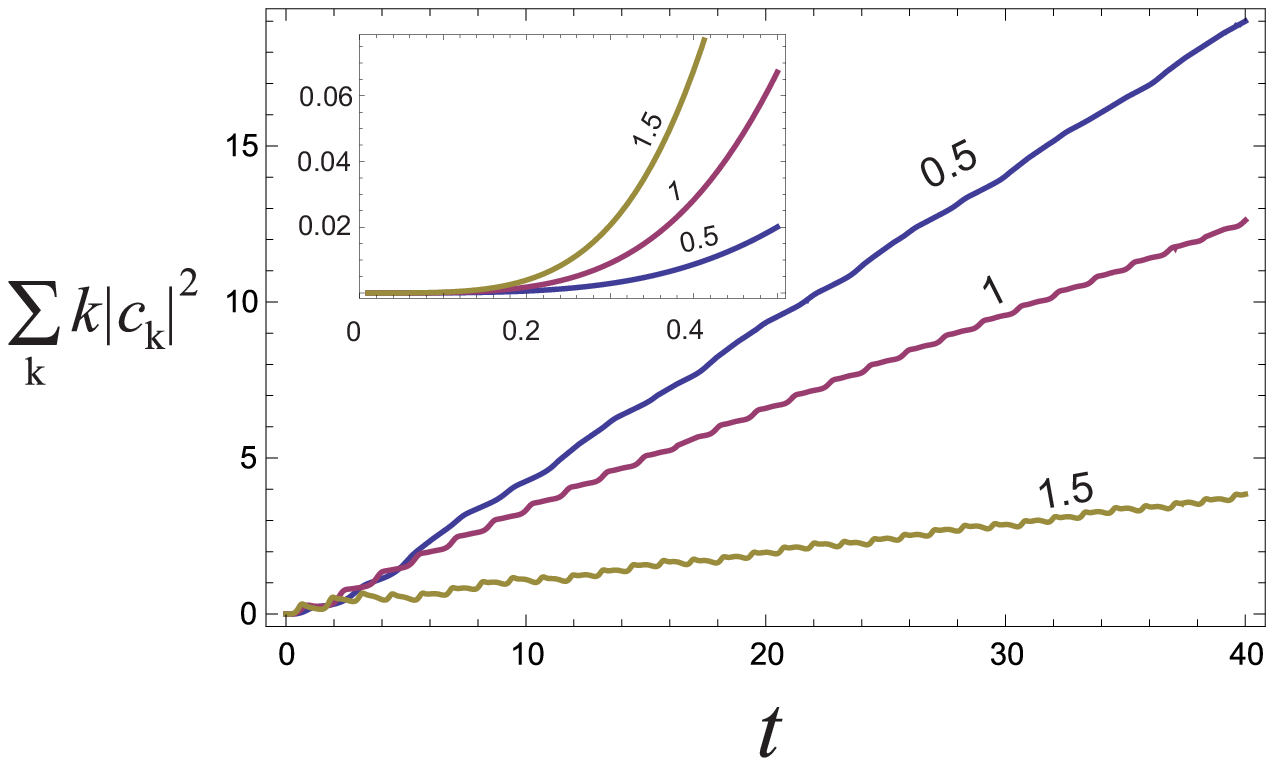}
\caption{(Color online) Numerical plot for the sum ${\displaystyle%
\sum\limits_{k=1}^{\infty }}k\left\vert c_{k}\right\vert ^{2}$ as function
of time for a Hamiltonian $H$ with spectral function $\protect\lambda \left(
\protect\theta \right) =(c-\cos \protect\theta )^{2}$ and for initial
Gaussian state $\protect\beta \left( \protect\theta \right) =1$ . The
top-most curve (blue) correpsonds to the critical Hamiltonian with $c=0.5$,
the middle curve (magenta) to the critical Hamiltonian with $c=1$, and the
lowest curve (yellow) to a gapped Hamiltonian with $c=1.5$. One clearly
recognizes a linear increase with time. The insert shows the quadratic short-time
evolution.}
\label{fig:entropy}
\end{figure}

In Fig.~\ref{fig:entropy} we have plotted the right hand side of eq.(\ref%
{LowerBoundFourier}) numerically evaluated for an initial state with
spectral function $\beta (\theta )=1$ and a Hamiltonian $H$ with spectral
function $\lambda (\theta )=(c-\cos (\theta ))^{2}$. If $c>1$, $H$ has a
finite excitation gap as there are no real zeroth of $\lambda (\theta )$. If
the gap vanishes, i.e. for $c\leq 1$ the ground state of $H$ becomes
critical. One clearly recognizes a linear increase with time in all cases.
That the presence of an excitation gap is irrelevant here is not surprising, as 
the initial state has a finite overlap
with excited states except in the trivial case where it coincides with
the ground state.

For short times one should expect that the sum grows quadratically in time
\cite{Unanyan2010}. The spectral function $\Lambda \left( \theta ,t\right) $
(\ref{symbolfunction}) for small $t$ scales approximately quadratic in $t$
\begin{equation}
\Lambda \left( \theta ,t\right) \approx \beta \left( \theta \right) \left[
1-\left( \beta ^{2}\left( \theta \right) -\lambda (\theta )\right) t^{2}%
\right] .  \label{shortTime}
\end{equation}%
The correction to the initial spectral function is proportional to
difference $\beta ^{2}\left( \theta \right) -\lambda (\theta )$, as was
expected. The Fourier coefficients $c_{k}$ (\ref{Fourier}) can then easily
be calculated $c_{k}\approx \xi _{k}+t^{2}\delta _{k}$ ,where $\xi _{k}$ and
$\delta _{k}$ are some constant numbers. From this one can calculate the
sum (\ref{LowerBoundFourier}) for short time  which yields
\begin{equation}
S\geq \varkappa _{1}+\varkappa _{2}t^{2}  \label{EntropyShortTime}
\end{equation}

In the following we will derive an analytic estimate for the lower bound to
the entropy for large times. Note that $\beta ^{2}(\theta )=\lambda (\theta )
$ corresponds to an initial state that is an eigenstate of $H$ and thus has
no time evolution at all. In any real system, the number of oscillators in
the chain is finite and therefore to neglect boundary effects in the
thermodynamic limit it is necessary consider time intervals $t\leq L/v$,
where $v$ is the speed for excitations after a quench, the so-called
Lieb-Robinson speed \cite{Lieb-Robinson} (see also: \cite{footnote}), and $L$
is the system size. For the sake of simplicity of the derivations
we consider a Hamiltonian $H$ with a finite excitation gap. The derivation
for a non-gapped Hamiltonian is more involved and will not be
presented here. As noted above the presence of a gap is however 
irrelevant. In the following we use an alternative
expression for ${\sum\limits_{k=1}^{\infty }}k\left\vert c_{k}\right\vert
^{2}$ which is very useful for numerical and analytical calculations. By
Parseval's theorem this sum can be rewritten as ${\displaystyle%
\sum\limits_{k=1}^{\infty }}k\left\vert c_{k}\right\vert ^{2}=\frac{1}{8\pi
^{2}}{\displaystyle\int\limits_{-\pi }^{\pi }}{\displaystyle\int\limits_{\pi
}^{\pi }}d\eta _{1}d\eta _{2}\frac{\ln ^{2}\frac{\Lambda \left( \eta
_{1}-\eta _{2},t\right) }{\Lambda \left( \eta _{1}+\eta _{2},t\right) }}{%
\sin ^{2}\eta _{2}}$. Making use of the inequality $\left\vert \ln
\left\vert \frac{x}{y}\right\vert \right\vert >\frac{1}{M}\Bigl\vert|x|-|y|%
\Bigr\vert,$ $0<\ \left\vert x\right\vert ,\left\vert y\right\vert \leq M$
one finds
\begin{equation}
S>\frac{1}{M^{2}}{\displaystyle\sum\limits_{k=1}^{\infty }}k\left\vert
b_{k}\right\vert ^{2},
\end{equation}%
where $M=\max \Lambda \left( \theta ,t\right) $, and
\begin{equation}
b_{k}=\frac{1}{2\pi }\int_{0}^{2\pi }d\theta \,\Lambda ^{-1}(\theta
,t)\,\exp \left( -i\theta k\right) .
\end{equation}%
The coefficients $b_{k}$ have a simple physical meaning: They determine the
correlations in momentum space over a distance $k$, i.e. $\langle \Psi
(t)|p_{i}p_{i+k}|\Psi (t)\rangle \sim b_{k}$. With this we have
\begin{equation}
S>\frac{1}{M^{2}}{\sum\limits_{k=1}^{\infty }}k\left( \varsigma _{k}+\mu
_{k}\left( t\right) \right) ^{2}  \label{LowerBound1}
\end{equation}%
where we have decomposed $\Lambda ^{-1}(\theta ,t)$ in a time independent
and a time dependent term
\begin{align}
\varsigma _{k}& =\frac{1}{4\pi }{\displaystyle\int\limits_{0}^{2\pi }}%
d\theta \frac{\lambda \left( \theta \right) +\beta ^{2}\left( \theta \right)
}{\beta \left( \theta \right) \lambda \left( \theta \right) }\,\cos \left(
k\theta \right) , \\
\mu _{k}\left( t\right) & =\frac{1}{4\pi }{\displaystyle\int\limits_{0}^{\pi
}}d\theta \frac{\lambda (\theta )-\beta ^{2}(\theta )}{\lambda (\theta
)\beta (\theta )}\cos \left( 2t\lambda ^{1/2}(\theta )\right) \cos \left(
k\theta \right) .  \label{correlationwithoutlog}
\end{align}%
The term proportional to $\varsigma_k ^{2}$ in eq.(\ref{LowerBound1}) does not
depend on time and can be disregarded. The second term can be rewritten
using the triangular inequality and Parseval theorem to give $\left\vert {%
\sum\limits_{k=1}^{\infty }}k\varsigma _{k}\mu _{k}\left( t\right)
\right\vert \leq \Bigl[ {{\sum\limits_{k=1}^{\infty }}k\varsigma _{k}^{2}{%
\sum\limits_{k=1}^{\infty }}k\mu _{k}^{2}\left( t\right) }\Bigr] ^{1/2}=%
\Bigl[ C_{1}{\displaystyle\sum\limits_{k=1}^{\infty }}k\mu _{k}^{2}\left(
t\right) \Bigr] ^{1/2}.$ The time dependence of this term is thus given by
the square root of the term containing $\mu _{k}(t)^{2}$.

We now show that the term $\sim \mu _{k}(t)^{2}$ in eq.(\ref{LowerBound1})
is bounded from below by a function linear in $t$. To this end we evaluate
the integral in eq.(\ref{correlationwithoutlog}) for $\mu _{k}(t)$ by the
method of stationary phase (the role of the large parameter is played by $t$%
). The stationary points of the phase $2\lambda ^{1/2}\left( \theta \right)
\pm \frac{k}{t}\theta $ are the solution of
\begin{equation}
\frac{1}{\lambda ^{1/2}\left( \theta ,t\right) }\frac{d\lambda \left( \theta
,t\right) }{d\theta }\pm \frac{k}{t}=0.  \label{stationary}
\end{equation}%
As the interaction matrix $V$ is of finite range, the spectral function $%
\lambda \left( \theta \right) $ is a trigonometric polynomial of finite
degree $K.$ By the theorem of Bernstein \cite{Szego} one has $\max
\left\vert \frac{d\lambda \left( \theta \right) }{d\theta }\right\vert \leq
K\max \lambda \left( \theta \right) $, and therefore if $k\geq k_{\mathrm{max%
}}=\frac{K\max \lambda \left( \theta \right) }{\sqrt{\min \lambda \left(
\theta \right) }}\,t=v_{g}\,t$ eq. (\ref{stationary}) has no real solutions.
For these values of $k$ all $\mu _{k}\left( t\right) $ are exponentially
small in agreement with the finite Lieb-Robinson speed.
On the other hand, when $t$ and $k$ are large but $k\leq v_{g}t$ one
finds
\begin{equation}
\mu _{k}\left( t\right) \approx \frac{1}{\sqrt{t}}{\displaystyle%
\sum\limits_{m=1}^{W}}F_{m}\left( \frac{k}{t}\right) \cos \left(
tG_{m}\left( \frac{k}{t}\right) +\varphi _{m}\right) ,
\label{stationaryPhase}
\end{equation}%
where $F_{m}\left( \frac{k}{t}\right) $ and $G_{m}\left( \frac{k}{t}\right) $
are some "nice" functions. $F_{m}\left( \frac{k}{t}\right) $ is proportional
to $\int \mathrm{d}\theta \left[ \lambda \left( \theta \right) -\beta
^{2}\left( \theta \right) \right] $ which quantifies the difference between
initial state and ground state of $H$. If both agree, i.e. if $\beta(\theta)^2
=\lambda(\theta)$, the coefficient vanishes.
The integer $W$ is the number of
stationary points, which must be finite because $\lambda \left( \theta
\right) $ is a trigonometric polynomial of finite degree. $\varphi _{m} =
\pm \frac{\pi }{4}$ depending of the sign of the
second derivative of $\lambda ^{1/2}\left( \theta \right) $ at the
stationary points. Thus ${\sum\limits_{k=1}^{\infty }}k\mu _{k}^{2}\left(
t\right) >{\sum\limits_{k=1}^{tv_{g}}}k\mu _{k}^{2}\left( t\right) \approx {%
\sum\limits_{k=1}^{v_{g}\,t}}\frac{k}{t}\left\vert {\sum\limits_{m=1}^{W}}%
F_{m}\left( \frac{k}{t}\right) \cos \left( tG_{m}\left( \frac{k}{t}\right)
+\varphi _{m}\right) \right\vert ^{2}$ By replacing summation by integration
and neglecting all highly oscillating terms we arrive at
\begin{equation}
{\sum\limits_{k=1}^{\infty }}k\mu _{k}^{2}\left( t\right) >\frac{t}{2}{%
\int\limits_{0}^{v_{g}}}\mathrm{d}x\,x{\sum\limits_{m=1}^{W}}F_{m}\left(
x\right) ^{2}=\alpha _{2}\,t+\,\mathcal{O}\left( t^{1/2}\right) .  \notag
\end{equation}%
Thus
\begin{equation}
S\,>\,C\,t,  \label{result}
\end{equation}%
with
\begin{equation}
C=\frac{1}{2M^{2}}\int_{0}^{v_{g}}\mathrm{d}xx\sum_{m=1}^{W}\Bigl(F_{m}(x)%
\Bigr)^{2}
\end{equation}%
being a finite time-independent constant. Eq.(\ref{result}) is the main
result of our paper. It constitutes a lower bound to the scaling of the
entanglement entropy $S_{\alpha },\alpha \leq 1$ with time in a
one-dimensional system of coupled harmonic oscillators. Eq.(\ref{result})
implies that the bond dimension of the matrices used in an MPS
representation needs to increase exponentially in time to allow for a
faithful representation of the dynamical many body wavefunction. This means
that in contrast to fermionic systems, where at least for certain initial
conditions a simulation of the long-time dynamics is possible, for harmonic
oscillator systems this is in general impossible. The discussion can be
extended to $d$ dimensions. In higher dimensions the form of the reduced
density matrix is the same as Eq.(\ref{reducedDensityMatrix}). To calculate
determinants of Toeplitz matrices one can apply the $d$-dimensional Szeg\"{o}
theorem \cite{Linnik}.

The authors would like to thank J. Eisert for stimulating discussions. The
financial support of the DFG through the SFB-TR49 is gratefully acknowledged.

\end{document}